# Band-edge Excitation of Carotenoids Removes S* Revealing Triplet-pair Contributions to the $S_1$ Absorption Spectrum.

Daniel W. Polak*[1], Andrew J. Musser[1], George A. Sutherland[3], Alexander Auty[2], Federico Branchi[4], Branislav Dzurnak[1], Jack Chidgey[3], Giulio Cerullo[4], C. Neil Hunter[3], Jenny Clark*[1]

[1] Department of Physics and Astronomy, University of Sheffield, Hounsfield Road, Sheffield S3 7RH, UK

[2] Department of Chemistry, University of Sheffield, Brook Hill, Sheffield S3 7HF, United Kingdom

[3] Department of Molecular Biology and Biotechnology, University of Sheffield, Sheffield S10 2TN, U.K

[4] IFN-CNR, Dipartimento di Fisica, Politecnicodi Milano, Piazza L. da Vinci 32, I-20133 Milan, Italy

*email: dwpolak1@sheffield.ac.uk, jenny.clark@sheffield.ac.uk

## Abstract:

The nature of the low-lying electronic states in carotenoids has been debated for decades. We use excitation-dependent transient absorption spectroscopy and comparison with published results on β-carotene to demonstrate that the so-called S* feature in astaxanthin, echinenone and spheroidenone spectra is due to an impurity in the sample. Excitation at the absorption band-edge results in a transient absorption spectrum dominated by the pure $S_1$ photo-induced absorption, with no contribution from impurities ('S*'). We find that this $S_1$ absorption resembles the triplet excited-state absorption spectrum, shifted by ~200meV. To explain this, we suggest that the individual triplets that make up the dominant triplet-pair (TT) configuration of $2A_g^-$ contribute strongly to the $S_1$ absorption band. Comparison with recent literature on molecular polycrystalline films which undergo singlet fission suggests the shift could be related to the binding energy of the triplet pair state. These findings have implications for understanding triplet-pair states in organic semiconductors and the excitation-energy dependence of singlet fission in carotenoid aggregates and biological systems.

## Introduction

Over 600 carotenoids exist in nature [1–3], covering a wide variety of functions from photo-protection in human vision, to energy harvesting in photosynthesis [4–10]. These roles are enabled by the complex properties of carotenoids where – assuming $C_{2h}$ symmetry – the lowest-lying singlet excited-state ($S_1$) is highly electron-correlated, described as a pair of triplets (TT) with overall $A_g$ symmetry (the $2A_g^-$ state) [11–22]. This state carries the same symmetry as the ground-state ($1A_g^-$), therefore $2A_g^-$ ($S_1$) has negligible oscillator strength. One-photon absorption is instead predominantly into the $1B_u^+$ singlet state, which is typically assumed to be the second singlet excited-state, $S_2$ [11–15,17–20]. Internal conversion from $1B_u^+$ to $2A_g^-$ occurs in <200fs [18,23], followed by non-radiative decay to the ground-state ($1A_g^-$) within 20ps [11–15,17–20]. This rapid non-radiative decay makes carotenoids perfect photo-protective materials for rapidly dissipating electronic energy into heat [4–9].

$S_1$ is a highly correlated 2-electron state, described as a bound pair of triplets. Similar (TT) states have been observed in acene [24–28] and heteroacene polycrystalline films [29], and are of current interest in the field of organic electronics. Such interest stems from the fact that triplet-pair states are of current interest in the field of organic electronics as the triplets can transfer charge or energy *independently*, either to an electron acceptor [30] or a triplet acceptor [31,32]. Thus, a single photon produces two excitations that could be harvested in solar cells, potentially increasing solar energy generation efficiency by 30% [33] and overcoming the Shockley-Queisser limit [34]. Interestingly, despite calculations showing the $2A_g^-$ state to have dominant $^1$(TT) character, to our knowledge no experimental comparison between triplet ($T_1$-$T_n$) and $2A_g^-$ excited-state absorption spectra has been presented.

The complex energy landscape in carotenoids has led to decades of controversy in the assignment of the excited-state processes [35]. The nature of the low-lying excited-states and internal conversion processes between them are particularly important for many biological processes and potentially for singlet fission applications. We present below a brief discussion of the current understanding of carotenoid excited-state energies and photophysics. For more in-depth discussion, we refer the reader to references [3,35].

In their seminal paper, Tavan and Schulten and later Schmidt and Tavan calculated the optically accessible states for polyenes [15,36]. Their calculated energies match estimated gas-phase energies remarkably well[36] and we plot their reported relaxed gas-phase energies[36] in Figure 1a.

In addition to $2A_g^-$ and $1B_u^+$ states, several other low lying excited-states were calculated[15,35], including the one-photon-inaccessible $1B_u^-$. Calculations of vertical gas-phase energies indicate that at conjugation lengths below N=6, the $1B_u^+$ state lies below the $1B_u^-$ state [36], while at longer conjugation lengths (N>6), these states switch order so that the $1B_u^-$ state becomes $S_2$. This state ordering suggests that – in the gas phase – both the $1Bu^+$ and the $1B_u^-$ states must be considered when describing excited-state decay pathways in carotenoids with longer conjugation lengths [35,37].

Experimentally, the $1B_u^-$ state was first invoked to explain the dependence of transition rate on energy gap for the $1B_u^+$ to $2A_g^-$ internal conversion that doesn't follow the expected exponential energy gap law [38,39]. Since then a raft of literature has suggested the involvement of $1B_u^-$ or other states in energy transfer [3], internal conversion [35,40], triplet generation [41,42], and charge-transfer[43] both in solution and embedded within a protein [3,15,35,44].

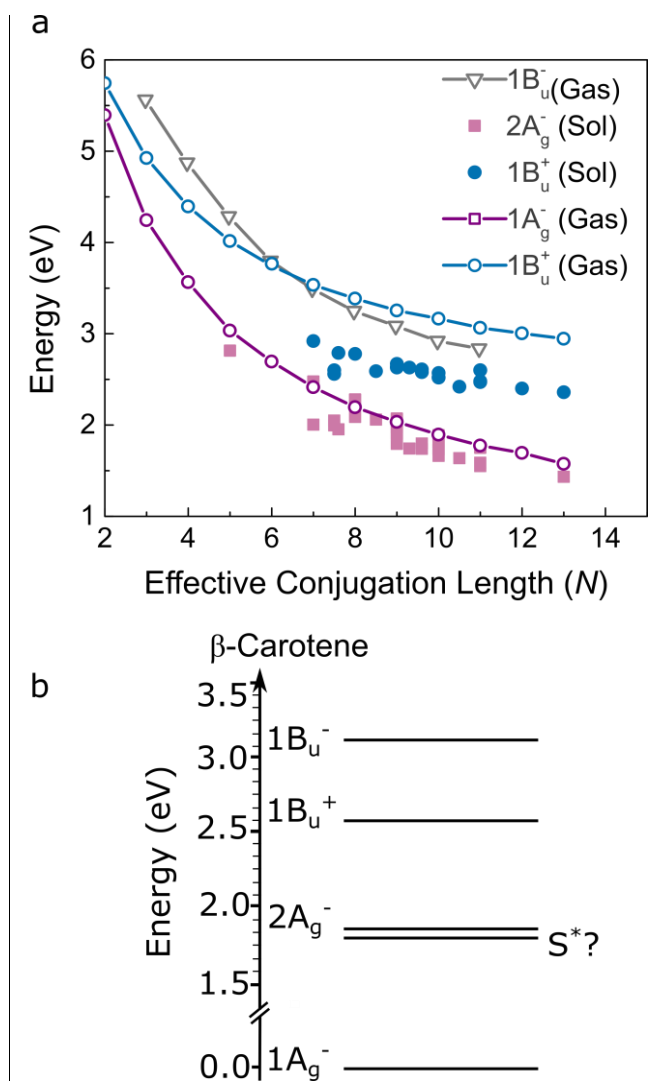

Figure 1: Energies of low-lying excited-states in carotenoids. (a) shows a summary of carotenoid energies from literature. Solid markers indicate measured solution phase energies reported in Ref [3], taking $1B_u^+$ energies from the ground state absorption spectrum, and $2A_g^-$ energies from emission [3,45–51]. Open circles indicate estimated gas phase energies reported in Ref [36]. $1B_u^+$ is stabilised by >0.5eV on transition from gas-phase to condensed-phase, while $2A_g^-$ doesn't shifts significantly. (b) shows the proposed energy levels of β-carotene taken from literature, including the 'controversial' S* state [35].

However, in condensed-phase environments such as solution or protein, the $1B_u^+$ state should be stabilised more than the covalent states ($1B_u^-$, $2A_g^-$) because of its large oscillator strength (suggested to be up to an order of magnitude larger [36]), see Figure 1 [3,36,45–55]. This stabilisation suggests that $1B_u^-$ should play no role in condensed-phase carotenoid photophysics. In other words, for all-*trans* carotenoids excited into the lowest-energy absorption band, calculations suggest the photophysics should depend on only three states: $1A_g^-$ ($S_0$), $2A_g^-$ ($S_1$) and $1B_u^+$ ($S_2$).

This 3-states model is supported by recently published work from Liebel *et al.* who used vibrational coherence spectroscopy to measure β-carotene in solution, showing that internal conversion occurs via a conical intersection and does not involve any other electronic states [19]. A conical intersection or avoided crossing between $1B_u^+$ and $2A_g^-$ would result in an internal conversion rate between $S_2$ and $S_1$ that does not depend exponentially on energetic gap, as is observed [38,56]. Indeed, at the ground-state geometry, $2A_g^-$ is calculated to lie higher in energy than $1B_u^+$ [21]. On reorganisation to the relaxed $2A_g^-$ geometry, it lies > 1eV below $1B_u^+$ [21,22]. This large geometric reorganisation is reflected in the large Huang-Rhys parameter required to fit the weak $2A_g^-$ emission spectra at low temperature [57,58].

One problem with the 3-state model comes from transient absorption measurements where a photo-induced absorption, to the blue wavelength side of the $S_1$-$S_n$ absorption spectrum, has been observed with lifetimes longer than the $2A_g^-$ state ($S_1$), Figure 3. In order to explain this feature, a new intermediate state called 'S*' has been proposed (Figure 1b). S* was first observed in light harvesting complex 1 containing sprillioxanthin [59,60], and was later suggested to be a precursor to triplet formation via singlet fission in these complexes [41,42]. A similar photo-induced absorption, with a pronounced excitation-energy dependence, was also measured in dilute solutions of β-carotene by Larsen *et al*. [37] and assigned to S*.

There is, however, growing evidence that 'S*' is not a separate excited-state. For example, Wohlleben *et al.*, investigated several carotenoids using pump-deplete-probe spectroscopy and concluded that the 'S*' feature is a hot ground state [40]. The pump-deplete-probe technique uses a third pulse between pump and probe to selectively depopulate the $1B_u^+$ state. It was shown that 'S*' is not depleted by the pulse while the $2A_g^-$ state is [40]. As a result the authors concluded that 'S*' cannot be populated in parallel with the $2A_g^-$ state [40] and instead suggested that the 'S*' feature is due to population of an excited vibrational level of the ground state [40]. To explain this, they suggest that the ground state is populated by either impulsive stimulated Raman scattering or non-radiative decay from the $2A_g^-$ state.

'S*' as a hot ground state was later disproved by comparing narrowband vs broadband excitation by Jailaubekov *et al.* [16]. Impulsive stimulated Raman scattering efficiency is expected to scale with bandwidth of the pump [16]. The authors demonstrate that 'S*' is not populated sequentially from $S_1$ and that, crucially, population of 'S*' does not depend on pump spectral breadth [16]. The latter implies that impulsive stimulated Raman scattering does not populate 'S*' [16]. Weerd *et al.* and others [60–63] instead suggest that a sub-population of carotenoids undergo a photo-induced change in conformation during relaxation from $1B_u^+$ ($S_2$) [60–63].

Recently Ostroumov *et al.* reported transient absorption measurements before and after purification of all-*trans*-β-carotene [64]. The long-lived component in the transient absorption spectra is entirely absent in the pure all-*trans* sample regardless of excitation conditions [64]. This suggests that the so-called 'S*' feature is due to a population of impurities in the sample [64]. The authors also collected spectra from each isolated impurity, finding similar ground-state absorption spectra to all-*trans*-β-carotene [64], suggesting the impurities are *cis* isomers [65]. The isomers are known to form over time in solution at room temperature, with a 20% population of a central carbon-carbon double bond *cis* isomer at equilibrium [66]. With the impurities removed, the entire transient absorption spectrum after 1ps can be described by a single transition ($S_1$-$S_n$). This is supported by work from Balevicius *et al.*, who were able to model the whole transient absorption spectrum of β-carotene (after purification) including only three states: $1A_g^-$, $2A_g^-$ and $1B_u^+$. [11,64].

Here we measure excitation-energy dependent transient absorption spectroscopy of a xanthophyll (astaxanthin), a linear carotenoid (spheroidenone) and a carotenoid with non-symmetric carbonyl groups (echinenone) and demonstrate that they follow the same trend as β-carotene, described previously. We are able to confirm that no intermediate states are required to describe the internal conversion from the $1B_u^+$ state to the ground state of these four carotenoids. Having removed the 'S*' contribution, we analyse the shape of the $S_1$ photo-induced absorption spectrum to demonstrate that it is consistent with the configuration of $S_1$ being dominated by a coupled triplet-pair, $^1(TT)$. We discuss this spectral behaviour in the context of literature describing singlet fission in acenes[24,25,27,67,68] and zethrenes[69].

## Methods:

β-carotene, astaxanthin and echinenone were purchased from Sigma-Aldrich at 95-97% purity and used as received. Spheroidenone was isolated from *Rhodobacter sphaeroides* as described in Chi *et al.* [70]. All carotenoids were dissolved in toluene at ~40µM for an OD of ~0.4-0.6 in 1mm path length cuvette. All transient absorption measurements were taken at room temperature with a setup similar to that described by Cerullo *et al.* [71,72]. A series of narrow band (15nm) pumps (under 1mW power) from 400 to 560nm and a visible probe pulse running from 500 – 700nm were supplied by home built non-collinear optical parametric amplifiers (NOPA). For the majority of results pump pulses were 100fs long, while for 'fast' TA a broadband (100nm) pump pulse centred at 500nm with 10-fs duration was used .

## Results and Discussion:

In Figure 2 we present ground state absorption spectra, energy levels and structures of β-carotene, astaxanthin, echinenone and spheroidenone. Each carotenoid has a strong absorption in the visible region due to the $S_0$-$S_2$ ($1B_u^+$) transition, with a series of red shifts between molecules [73] as the effective conjugation length increases from β-carotene (9.6) to astaxanthin (10.5) [1,3]. In addition, moving from β-carotene to astaxanthin and spheroidenone causes a pronounced loss of the vibronic structure in the ground state absorption due to the presence of carbonyl groups that adopt a range of conformations in polar environments, causing smearing of the vibronic peaks [73]. The additional absorption peaks around 375-425nm in spheroidenone are commonly assigned to *cis* isomers [62,73].

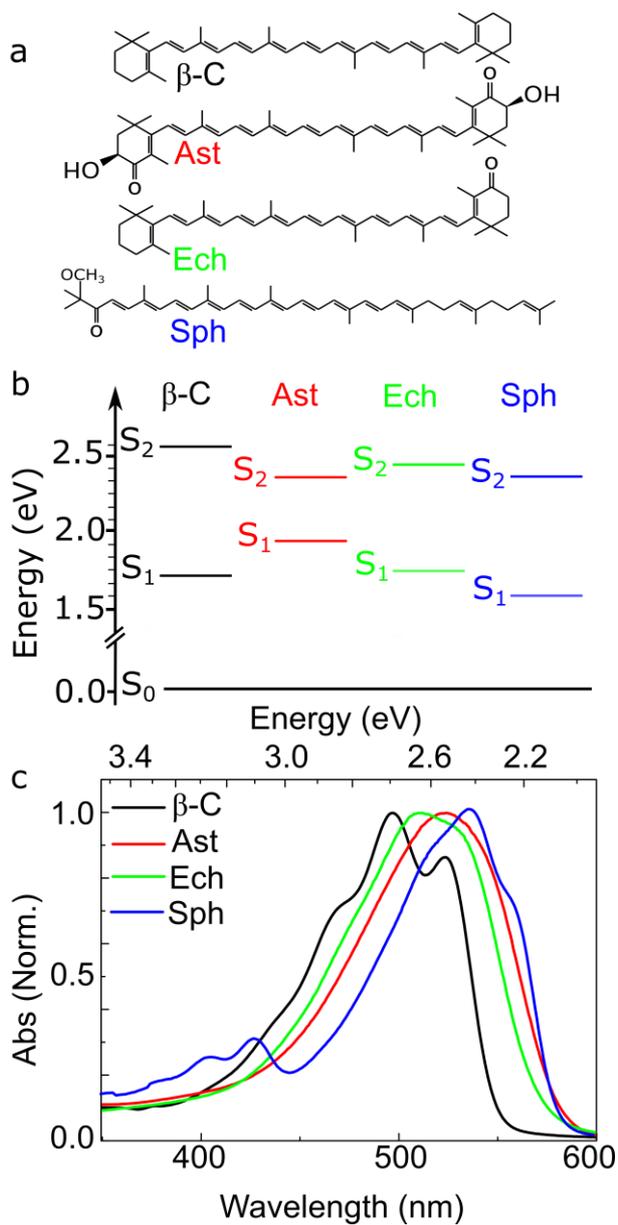

Figure 2: Carotenoid structures and energy levels. (a) Structures of β-carotene, astaxanthin, echinenone and spheroidenone. (b) Measured state energies relative to the ground state. $S_2$ and $S_1$ energies are taken from literature by Bachilo et al. [74], Musser et al. [75], Polivka et al. [3] and Cong et al. [62] for β-carotene, astaxanthin, echinenone and spheroidenone, respectively. (c) ground state absorption spectrum of each molecule in dilute (~40μM) toluene solution.

## 'S*' feature in transient absorption spectra is due to an impurity.

Figure 3 shows transient absorption spectra and kinetics for β-carotene and astaxanthin excited low (520nm) and high (400nm) in the absorption band. The photo-induced absorption peaked at 590nm (β-carotene) and 630nm (astaxanthin) is due to the $S_1$-$S_n$ transition [75,76]. This feature decays with a time constant of 10ps and 5ps for β-carotene and astaxanthin respectively, in agreement with literature [64,75]. With high excitation energy (400nm), a shoulder on the blue side of the $S_1$-$S_n$ peak, consistent with features assigned to S* in β-carotene, appears at 560nm and 540nm for β-carotene and astaxanthin, respectively [11,64].

This 'S*' peak decays slower than the rest of the spectrum, with time constants of 80ps (β-carotene) and 69ps (astaxanthin) Figure 3. The slow decay component is usually used as proof for the assignment of S* [16,37,40,61–63,77]. However, as discussed above, Ostroumov et al. discovered that a series of impurities that form in solution are responsible for this feature in β-carotene [64]. To reproduce these results, we measured the transient absorption spectra of β-carotene and astaxanthin excited across their absorption band (400-520nm) (Figure 3). We compare the excitation photon energy dependence of the peak intensity of the transient absorption feature usually assigned as S* to the absorption of the impurity identified by Ostroumov et al. [64]. The impurity absorption spectrum (Figure 3c, f, blue circles) was obtained by subtracting the all-*trans* absorption spectrum from the total sample spectrum from Ref [64], leaving only the collective absorption of the *cis* isomers [64]. For both β-carotene (Figure 3c) and astaxanthin (Figure 3f) we find a striking similarity between the two plots, with a complete loss of the impurity signal for excitation below 2.6eV (480nm) (Figure 3). The lower-energy onset of 'S*' absorption in astaxanthin is a result of the red-shift in absorption compared to β-carotene. We conclude that in both β-carotene and astaxanthin the long-lived feature assigned by others to S* is the impurity feature observed in Ref [64]. Interestingly, there is significantly less impurity (i.e. *cis* isomer) population in astaxanthin ($\frac{S^*}{S_1} = 0.045$) as compared to β-carotene ($\frac{S^*}{S_1} = 0.25$). A possible explanation for this is that astaxanthin has a much higher activation energy of isomer formation, found experimentally to be 101kJ/mol compared to 14kJ/mol for β-carotene [78,79].

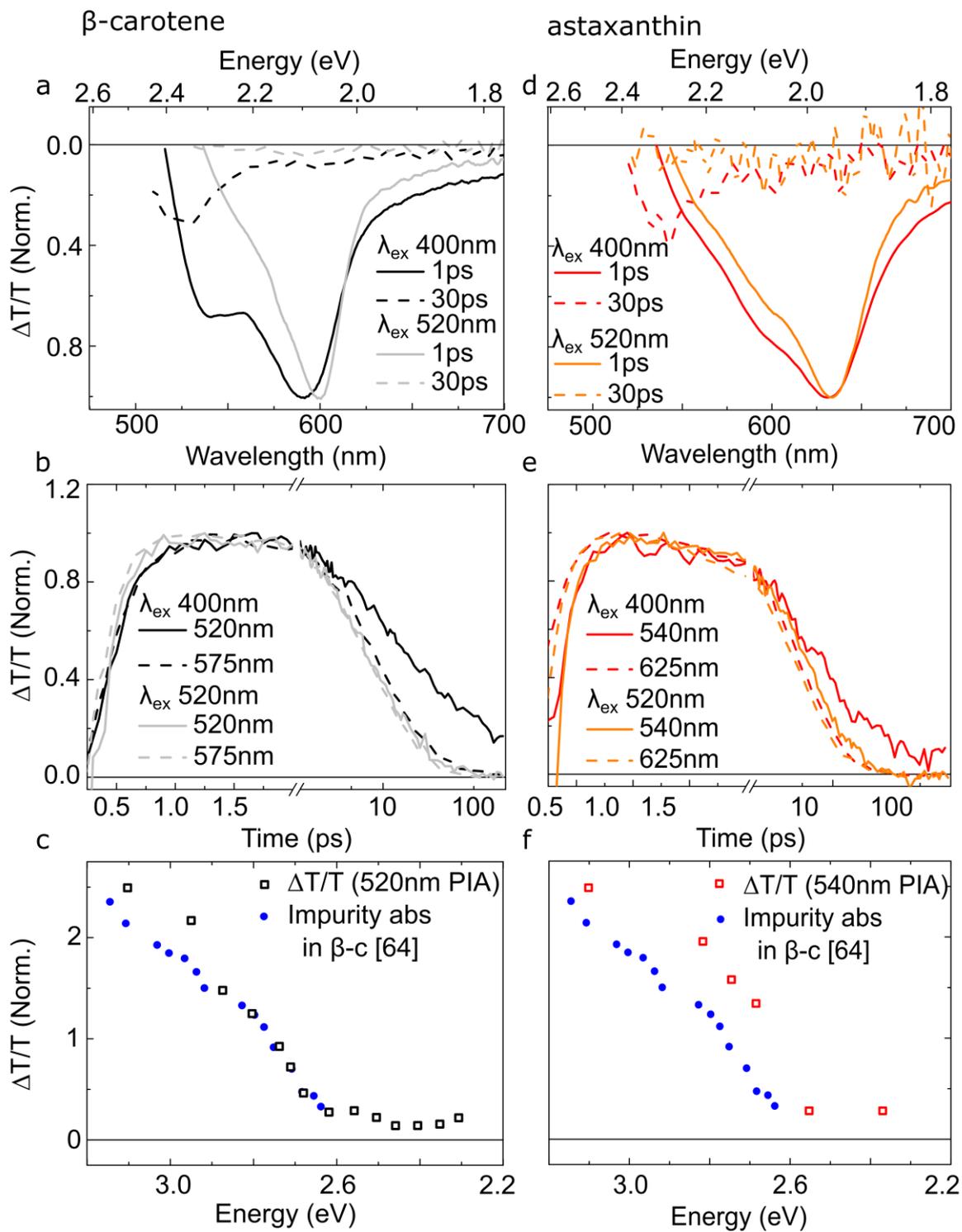

Figure 3: Excitation photon energy dependent transient absorption (β-carotene and astaxanthin). Transient absorption spectra (a,d) and kinetics (b,e) are shown with 400nm and 520nm excitation for β-carotene and astaxanthin respectively. (c) and (f) show the excitation-dependent peak intensity of the putative S* absorption feature in β-carotene (c) and astaxanthin (f) (Open markers). Also included is the ground state absorption of the impurity identified by Ostroumov *et al.* [64] (closed markers).

Figure 4 compares high time resolution transient absorption data for β-carotene and astaxanthin excited with broadband 10-fs pulses (2.21-2.57eV) versus narrowband excitation data shown previously (Figure 3). This pump overlaps with the carotenoid absorption spectrum without exciting above the 2480nm threshold for impurity excitation (see Figure 3c,f). It was shown by Jailaubekov *et al.* that narrowband (392-400nm) versus broadband excitation (389-416nm) has little effect on the proposed S* state in β-carotene [16]. We use a broader pump pulses range for both β-carotene and astaxanthin and find no evidence of 'S*' (Figure 4), confirming that a hot ground state is not required to explain the transient absorption data. In fact, the dynamics in Figure 4 show an identical rise and decay at the wavelengths traditionally assigned to 'S*' and $S_1$-$S_n$, suggesting that a single transition can fully explain the transient absorption spectrum of β-carotene and astaxanthin. In agreement with this assignment, Balevicius *et al.*, have shown that the transient absorption spectrum is consistent with a vibronic progression from a single electronic state [11]. We assign this to $S_1$-$S_n$.

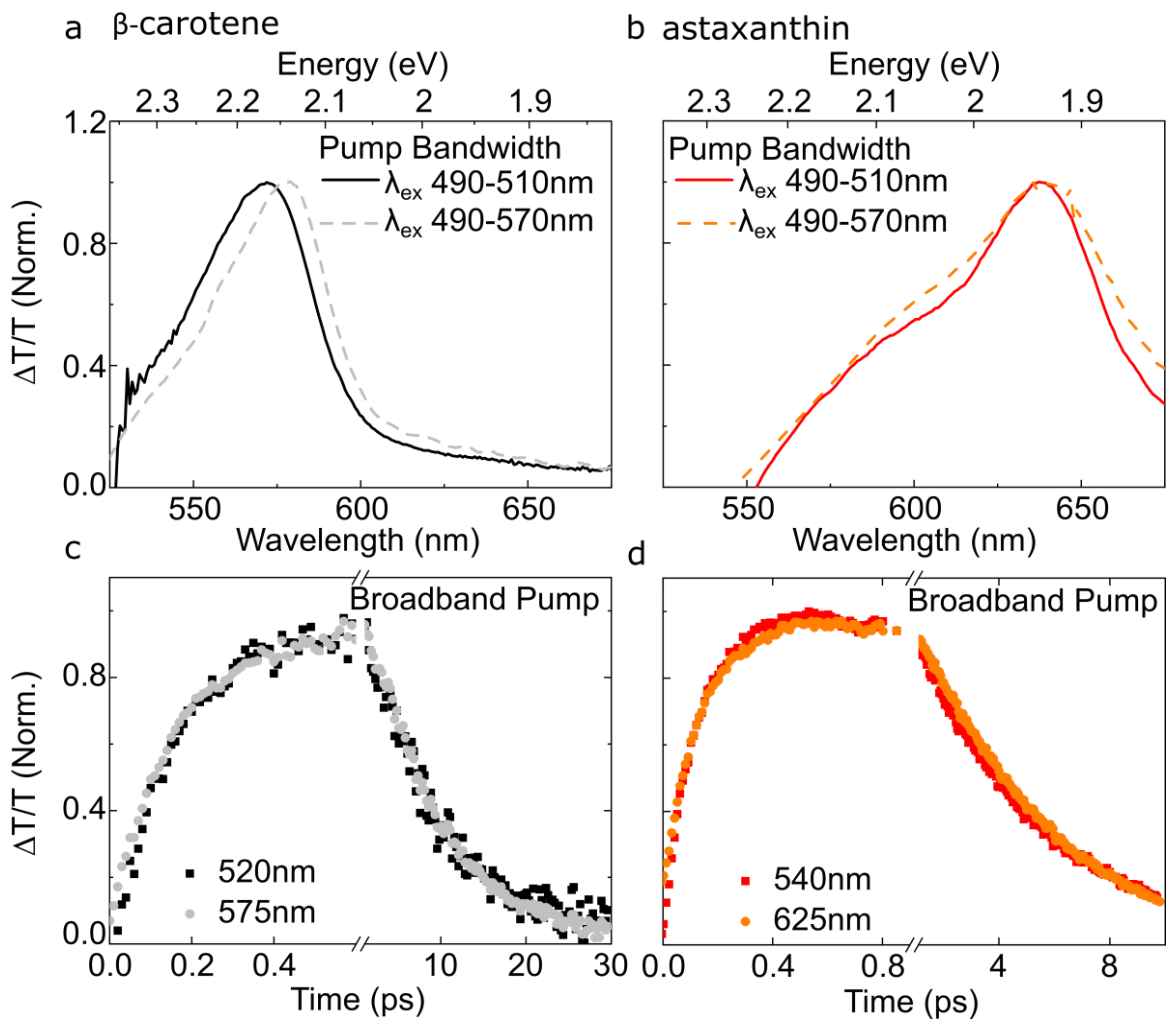

Figure 4: Narrowband versus broadband excitation transient absorption of carotenoids. Transient absorption spectra are shown at 1ps delay after narrowband (~490-510nm, solid) and broadband (~490-570nm, dashed) for β-carotene (a) and astaxanthin (b). High resolution transient absorption dynamics taken at the peak intensities for the two features associated with the $S_1$-$S_n$ transition for β-carotene (c) and astaxanthin (d).

Figure 5 shows equivalent data for echinenone and spheroidenone, excited with narrowband pulses above and below the onset energy for impurity absorption. A peak at 610nm (6.4ps lifetime) is measured for echinenone and a peak at 585nm (6.7ps lifetime) is measured for spheroidenone matching values for the $S_1$-$S_n$ transition from literature [73,80,81]. The small side peak visible at ~650nm for spheroidenone has been suggested to be due to the charge-transfer state character of the $S_1$ state, although this is contested [73,81].

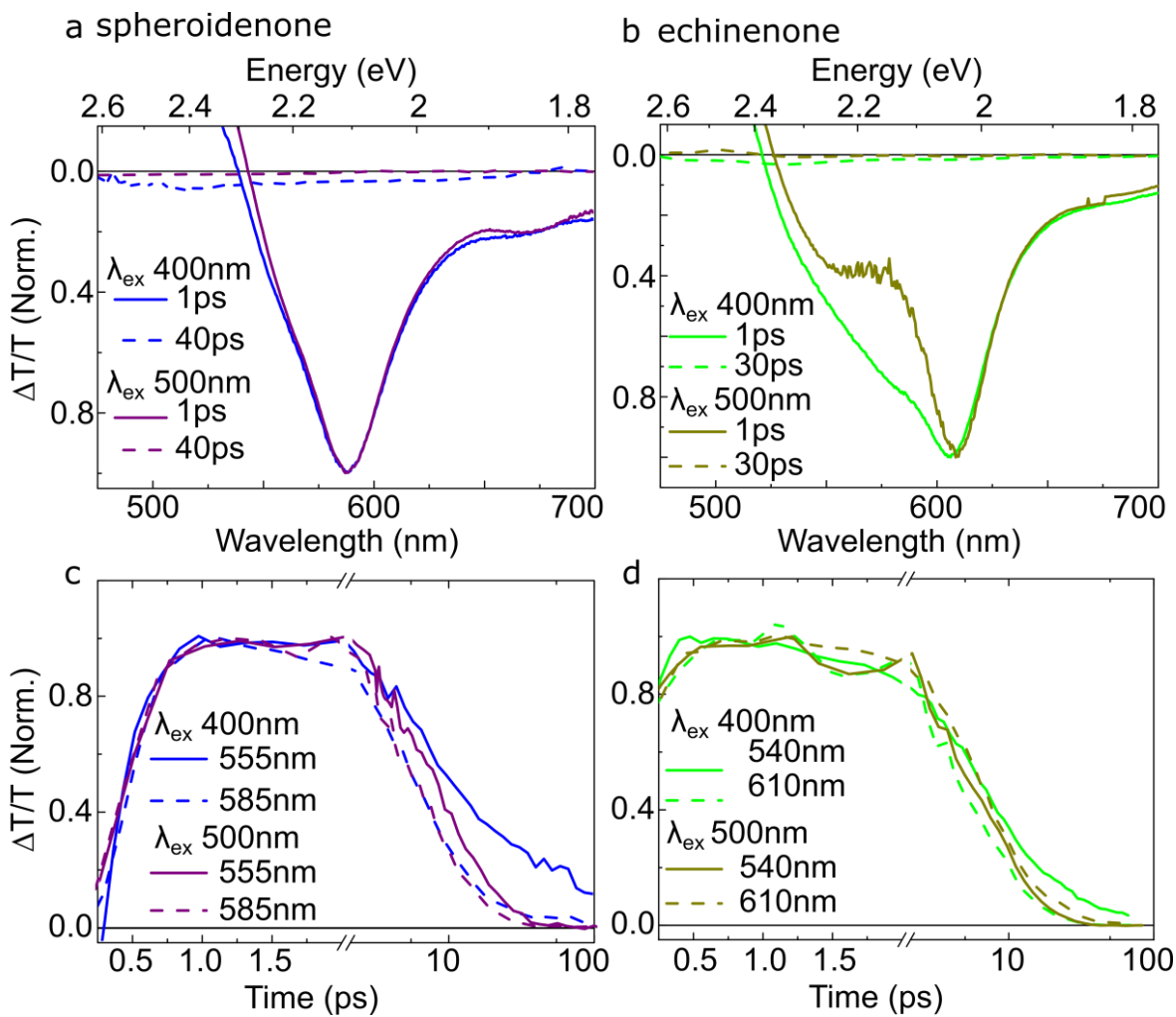

Figure 5: Excitation photon energy dependent transient absorption (spheroidenone and echinenone). Transient absorption data excited above and below the impurity excitation energy (400nm vs 500nm) for spheroidenone (a,c) and echinenone (b,d). Spectra are shown at two pump-probe delays (1ps and 30ps) for spheroidenone (a) and echinenone (b). Also included are dynamics taken at peak intensity for the impurity and $S_1$-$S_n$ transition for spheroidenone (c) and echinenone (d).

When comparing spectra excited above and below the onset absorption energy of the impurity identified above, we see a new feature appear with a lifetime of 33ps and 250ps in echinenone and spheroidenone respectively (Figure 5). In analogy with β-carotene and astaxanthin above we suggest that the longer-lifetime component is produced by a separate impurity population and is not a sign of a new electronic state. Indeed, with low-energy excitation the entire spectrum decays with the same dynamics, requiring no need to invoke

'S*'. We conclude that the $S_1$-$S_n$ transition is the only feature present in the transient absorption spectra with band-edge excitation of β-carotene, astaxanthin, echinenone and spheroidenone.

### The 2Ag- state is a (TT) state and its absorption spectrum can be described as a T-T* transition.

We now turn to examine the spectral shape of the band-edge-excitation transient absorption spectra of $S_1$ ($2A_g^-$) presented in Figures 3 and 5. Balevicius *et al.*, have shown that the transient absorption spectrum for β-carotene is consistent with a vibronic progression from a single electronic state, assigned to $S_1$ [11].

To understand the spectrum in more detail, we note that the dominant configuration in the wavefunction of $2A_g^-$ is a pair of triplets $^1$(TT), with other contributions from charge-transfer and e-h particle excitations [14,15,36]. Recent work by Khan and Mazumdar [82] suggests that a triplet-pair $^1$(TT) should absorb with a similar spectral shape to a triplet exciton ($T_1$). Barford *et al.* likewise suggest that the visible photo-induced absorption in linear polyenes is a T-T* transition[21]. More recently, empirical work by Lukman *et al.,* [69] suggests that the $^1$(TT) absorption should be blue-shifted compared with the isolated triplet $T_1$-$T_n$ photoinduced absorption spectrum by an energy proportional to the triplet-pair binding energy $E_b = 2E_T - E_{TT}$ [69]. In carotenoids, as the triplets which make up $2A_g^-$ are on the same chain, they should be strongly exchange-coupled and bound [83]. We estimate the binding energy for the β-carotene $^1$(TT) state (i.e. $2A_g^-$) from the difference between $2A_g^-$ (1.75eV ± 25meV [57]) and 2x$T_1$ (1.88eV ± 80meV [3,57,84]) to be 130 ± 105meV. With such strong binding between triplets, $^1$(TT) is expected to be a pure spin-singlet state [85–87]. As a singlet state, it does not require a spin flip to return to the ground state, with its lifetime is dominated by the standard gap-law for non-radiative singlet transitions [88].

In figure 6 we present a comparison between the transient absorption features of the $S_1$ state and a sensitized triplet spectrum ($T_1$-$T_n$) from literature [74,75,89]. Shifting the $S_1$-$S_n$ absorption spectrum by 190meV to higher energies for β-carotene (dashed) demonstrates a striking similarity between the spectral absorption shapes of $S_1$ and $T_1$. This suggests that, as expected, the $2A_g^-$ state can be described as a bound triplet-pair with triplet-like photo-induced absorption. A similar shift is visible in astaxanthin and echinenone, however the spectral shape diverges on the high-energy edge due to overlap with the ground state bleach. Interestingly, as expected from the empirical relationship derived in Ref.[69], the photoinduced absorption shift (190meV for β-carotene) is roughly proportional to the triplet-pair binding energy (130meV ± 105meV).

Although the magnitude of the photoinduced absorption shift matches the empirical relationship demonstrated in Ref.[69], the direction of the shift does not. In carotenoids, the $2A_g^-$ photo-induced absorption is shifted to lower energies compared with the $T_1$ photo-induced absorption. Assuming interaction between triplets does not change the $T_n$ energy ($E_{T_n}$), we would expect the $2A_g^-$ photo-induced absorption to be shifted by 130 ± 105meV to *higher* energy [69], instead of lower energy as we observe. This suggests the photo-induced absorption spectral shift depends not only on the $^1(TT)$ binding energy in carotenoids, but could stem from coupling to particle-hole excitations[90,91], charge transfer excitons[22] or conformational reorganisation[21,22].

The fact that $2A_g^-$ can be described as $^1(TT)$ raises further questions as to the classification of the transition from the bright singlet ($1B_u^+$) state to the dark multi-exciton ($2A_g^-$) state. To date this transition has been considered 'internal conversion' as it involves the transition between two singlet states. Singlet fission, on the other-hand, was defined as the formation of two 'free' (i.e. non-bound) triplets[83,92]. However, recent studies of acene and heteroacene materials[24–27,67,93–95] show that singlet fission creates a strongly exchange coupled (bound) pair of triplets that is similar to the $2A_g^-$ state described here. It is therefore reasonable to suggest that transition between $1B_u^+$ and $2A_g^-$ could also be described as 'intramolecular singlet fission' using the new definition of singlet fission in the acene literature.

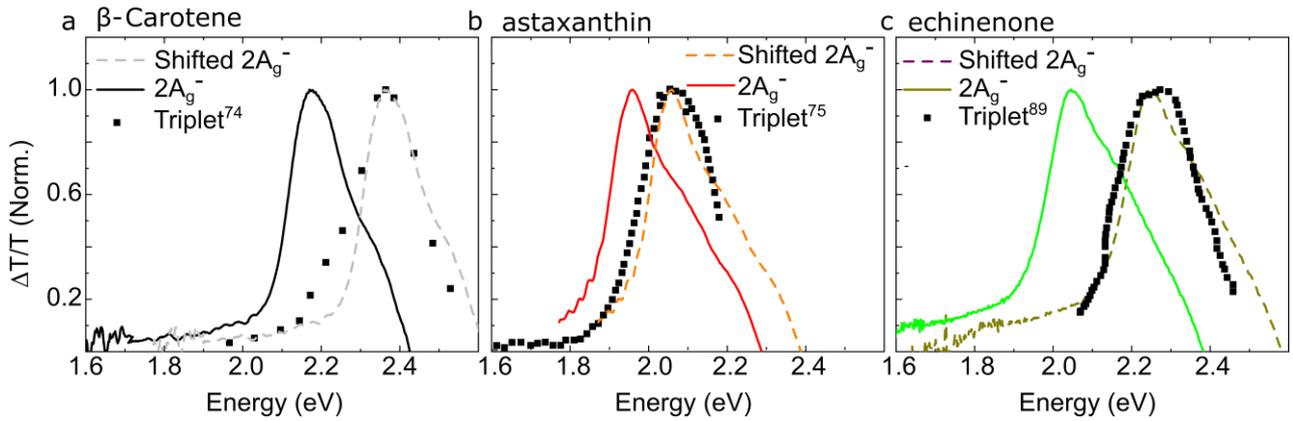

Figure 6: $S_1$ absorption is similar to $T_1$-$T_n$. $S_1$-$S_n$ and $T_1$-$T_n$ transitions for (a) β-carotene, (b) astaxanthin and (c) echinenone, dashed lines show the $S_1$-$S_n$ photo-induced absorption spectrum shifted by (a) 190meV, (b) 100meV and (c) 220meV to higher energies. $T_1$-$T_n$ spectra were taken from triplet sensitization from literature, β-carotene [74], astaxanthin [75], echinenone [89].

## Conclusion:

Here we have used excitation-energy dependent transient absorption spectroscopy of β-carotene, astaxanthin, echinenone and spheroidenone in comparison with published data on β-carotene[64] to show that the so-called 'S*' feature in all four carotenoids is due to impurities in the sample. Fortunately, with band-edge excitation the 'S*' feature vanishes leaving a pure $S_1$ photo-induced absorption spectrum. The shape of this spectrum in the visible spectral range can be explained as due to a coupled pair of triplets (TT), which is known to be the dominant configuration of $2A_g^-$ [15,21,22,36]. The excited-state absorption spectrum of $2A_g^-$ shifts by ~200meV for β-carotene, ~100meV for astaxanthin and ~200meV for echinenone compared with the absorption spectrum of $T_1$. These findings demonstrate that the transient absorption spectrum of carotenoids contains a wealth of information not yet understood. For example, the shift of the photo-induced absorption spectrum to lower energies when going from 'free' to 'bound' triplets is in the opposite direction as expected from similar studies on singlet-fission based molecular thin-films. Nevertheless, our understanding of excitation-dependent transient absorption spectroscopy of carotenoids, and the role of impurities, will help future investigations on excitation-dependent singlet fission in carotenoid aggregates and improve understanding of the role of carotenoids in nature.

Acknowledgments:

This work was made possible by contributions from the Lord Porter Laser Facility (EP/L022613/1), EPSRC Hybrid Polaritons (EP/M025530/1), Singlet fission in polyenes (EP/N014022/1), h2020 Laser Lab Europe (654148), BBSRC (BB/M000265/1), ERC advanced grant (SYNTHPHOTO) and the University of Sheffield for the Graduate Teaching Assistant scholarship and Vice Chancellor fellowship.